\documentclass[12pt,preprint]{aastex}

\shorttitle{On the Enhanced X-ray emission from SGRs}
\shortauthors{Ertan $\&$ Alpar 2003}

\begin{document}


\title{On the Enhanced X-ray Emission From SGR 1900+14 after the \\
    August $27^{th}$ Giant Flare}


\author{\"{U}. Ertan\altaffilmark{1} and 
M. A. Alpar\altaffilmark{2}}
\affil{Sabanc{\i} University, Orhanl{\i}$-$Tuzla 34956 
{\.I}stanbul/ Turkey}
 
\altaffiltext{1}{unal@sabanciuniv.edu}
\altaffiltext{2}{alpar@sabanciuniv.edu} 

\begin{abstract}

We show that the giant flares of soft gamma ray repeaters 
($E \sim 10^{44}$ erg)
can push the inner regions
of a fall-back disk out to larger radii by 
radiation pressure, while matter remains bound to the 
system for plausible parameters. 
The subsequent relaxation of this pushed-back matter can 
account for the observed enhanced X-ray emission after 
the August 27$^{th}$ giant flare of SGR 1900+14. 
Based on the results of our models, we estimate that the  
ratio of the fluences of the enhanced X-ray emissions to that of the 
preceding bursts remains constant for a particular SGR  with similar 
pre burst inner disk conditions, which is consistent with the four
different burst observations of SGR 1900+14.

\end{abstract}

\def\la{\raise.5ex\hbox{$<$}\kern-.8em\lower 1mm\hbox{$\sim$}}
\def\ga{\raise.5ex\hbox{$>$}\kern-.8em\lower 1mm\hbox{$\sim$}}
\def\be{\begin{equation}}
\def\ee{\end{equation}}  
\def\ba{\begin{eqnarray}}
\def\ea{\end{eqnarray}}  
\def\be{\begin{equation}}
\def\ee{\end{equation}}  
\def\ba{\begin{eqnarray}}
\def\ea{\end{eqnarray}}  
\def\m{\mbox}
\def\d{\partial}
\def\R{\right}
\def\L{\left}
\def\a{\alpha}
\def\Mdot*{\dot{M}_*}
\def\Mdotin{\dot{M}_{\mbox{in}}}
\def\Lin{L_{\mbox{in}}}
\def\Rin{R_{\mbox{in}}}
\def\Rout{R_{\mbox{out}}}
\def\Ldisk{L_{\mbox{disk}}}
\def\dEb{\delta E_{\mbox{burst}}}
\def\dEx{\delta E_{\mbox{x}}}
\def\Bb{\beta_{\mbox{b}}}
\def\Be{\beta_{\mbox{e}}}
\def\dMin{\delta M_{\mbox{in}}}
\def\dM*{\delta M_*}




\section{Introduction}

Soft gamma ray repeaters (SGRs) are neutron stars that emit short
($\la $ 1 s) and luminous ($\la 10^{42}$ erg s$^{-1}$) soft gamma ray
bursts in their active phases. The burst repetition time scales extend
from a second to years (see Hurley 2000 for a review). 
In their quiescent states, they emit persistent 
X-rays at luminosities similar to those of anomalous X-ray pulsars (AXPs) 
($L_{\m{}}\sim 10^{34} - 10^{36}$ erg s$^{-1}$). The spin periods of both
SGRs and AXPs are in a remarkably narrow range ($P \sim $ 5 - 12 s) 
(see Mereghetti 2000 for a review of AXPs). Four  SGRs 
(and one candidate) and
six AXPs are known up to date. Some of them were reported to be  
associated with supernova remnants indicating that they are young objects.
Recently, some AXPs also showed bursts
similar to those of SGRs, which probably imply that they belong
to the same class of objects.        

Over the burst history of SGRs, two giant flares were exhibited by SGR
0526-66 (Mazets et al. 1979) and SGR 1900+14 (Hurley et al. 1999). 
These giant flares are characterized by an initial hard spike with a peak
luminosity $\sim 10^{44}-10^{45}$ erg s$^{-1}$ which lasts a fraction of a
second. The hard spike is followed by an oscillating tail that
decays in a few minutes. Assuming isotropic emission  the fluence of
the entire giant flare is about $\sim 10^{44}$ ergs 
(Hurley et al. 1999; Feroci et al. 2001; Mazets et al. 1999). 

Magnetar models can explain the super-Eddington luminosities of the normal
and the giant bursts of SGRs by the sudden release of the very high
magnetic energies from inside the neutron stars 
(Thompson $\&$ Duncan 1995). 
In an alternative class of models,  
fall-back disks around young neutron stars 
can account for the period evolution of these systems, and in particular
for the period clustering of SGRs and AXPs (Chatterjee, Hernquist, 
$\&$ Narayan 2000; Alpar 2001).  
Thompson et al. (2000) argued that the high luminosity of a giant
flare would excavate any
accretion disk to a large radius (due to the radiation momentum) and
rebuilding of the entire disk takes months to years; so that the     
enhancement and the decay of the persistent X-ray flux after the giant   
flare could not be related to any disk accretion phenomenon. 

The persistent X-ray emission from SGR 1900+14 was  reported to
increase by a factor $\sim$ 700 about 1000 s after the giant flare. The
subsequent  decay is a power law with an index $\sim 0.7$ (Woods et
al. 2001). This increase and decay in the persistent
X-ray
emission of the SGR 1900+14 is our main interest here. 
It was proposed that 
the enhanced X-ray emission is due to the cooling of the
neutron star crust after being heated by the energy of the giant flare 
(Lyubarsky, Eichler, $\&$ Thompson 2002). Here  
we show by means of a numerical disk
model that $(i)$ the X-ray enhancement can be explained in terms of 
the viscous relaxation of a disk pushed back by the giant flare, 
$(ii)$ the amount of disk matter pushed out while remaining bound 
corresponds to a plausible fraction of the flare energy.
The origin of the giant flare, which is probably the release of the
high magnetic energy inside the NS by an instability, is not addressed
in our model.

In the next section, we summarize the X-ray observations of SGR 1900+14
revealing the large flux changes in the persistent X-ray emission following
the August 27 giant flare. In Sec.3, we present the details of the 
numerical disk models. The results of the model fits are discussed in 
Sec. 4. The conclusions are summarized in Sec. 5.

\section{The X-Ray Data}

The X-ray (2-10 keV) flux data following the August 27 giant flare of
SGR 1900+14 shown in Figures 1-3 was taken from Woods et al. (2001). 
The
first and the second data points (filled squares) are from RXTE/ASM
measurements, and correspond to $\sim$ 24 minutes and $\sim$ 2 hours 
after the giant flare (Remillard et al. 1998; see Woods et al. 2001 
for the measurements and the associated uncertainties). The 
two data points about 20 days after the giant flare are from the net
source intensity measurements of BeppoSAX (filled triangle) and ASCA
(filled circle) satellites. The remaining data points (crosses) are
estimated from the pulsed intensity measurements by RXTE/PCA (Woods
et al. 2001) as follows. Four BeppoSAX NFI observations of SGR 1900+14
(2000 March/April, 1998 September, and 1997 May) give similar pulsed 
fractions ($\sim 0.1$) despite the varying intensity, pulse profile,
and burst activity. In the light of these
observations, Woods et al.(2001) estimated the total source intensity
from the pulsed intensity measurements by assuming a constant pulsed
fraction (${\cal F}_{\m{rms}} \sim 0.11)$. There is a good agreement
between
these estimates, and the BeppoSAX and ASCA source intensity
measurements about 20 days after the giant flare (see Woods et al. 2001
for more details). The reported
relative pulsed fraction changes  along the  X-ray tails following the 
other observed bursts of SGR 1900+14 do not exceed a factor $\sim 2$
in the extreme case (Lenters et al. 2003). This does not affect the quality of
our fits, but might require a small modification of the model parameters 
presented here. 
 Keeping these uncertainties in mind we adopt for our
numerical model the X-ray data set shown in Figures 1-3, obtained by 
scaling with the pulsed fraction ${\cal F}=0.1$ where only the pulsed signal is
observed.  

\section{The Numerical Model}

Assuming isotropic emission,  
the total emitted energy during the giant flare is $\sim 10^{44}$
ergs (Mazets et al. 1999).  
A fraction of this emission is expected to be absorbed by the disk 
depending on the solid angle provided by the disk for the isotropic  
emission. For such a point-like emission at the center of the  
disk, the radiation pressure is expected to affect mostly the 
inner regions of the disk by pushing the inner disk matter to 
larger radii depending on the energy imparted to the disk matter.
This leads to large density gradients at the inner rim of the disk
immediately after the giant flare. We test whether the consequent  
viscous evolution of the disk can reproduce the X-ray flux data,
consistently with the 
reported energy arguments of the giant flare. 

In our model, we represent pushed-back inner disk matter, which we assumed 
to be formed by the radiation pressure of the giant flare, 
by a Gaussian surface density distribution 
$\Sigma(R,t=0) = \Sigma_{\m{max}}~ exp \left[-\left(\frac{R-R_0}{\Delta R}
\right)^2 \right]$, representing the pile up, 
added to the inner edge, at $R_0$,  
of the extended disk profile for which we chose the form  
$\Sigma = \Sigma_0 (R_0/R)$. 
$\Sigma_0$ is a
constant much less than $\Sigma_{\m{max}}$, $R$ is the radial distance
from
the center of the disk, and $R_0$ is the initial radial position of the 
center of the Gaussian.
This form of the extended disk is close to the surface density profile of
a standard thin disk (Shakura $\&$ Sunyaev 1973). 
In addition to the post-flare radius $R_0$, 
$\Sigma_{0}$, the Gaussian width and the maximum initial surface density 
$\Sigma_{\m{max}}$ 
(at the center of the Gaussian) are the free parameters of our model. 
The disk's inner radius $R_{\m{in}}$ (where the subsequent
inflow of the pushed-back  matter will be stopped by the magnetic
pressure), and the outer disk radius $R_{\m{out}}$  are  
kept constant throughout the calculations. 
A constant outer disk radius was chosen due to numerical
reasons. Outer disk properties can only affect the
inflow rate through the inner disk after several weeks or more in the
absence of large surface density gradients at the outer
disk regions. We 
use the one dimensional disk code described in Ertan $\&$ Alpar (2002),
originally constructed to simulate the black hole soft X-ray transient
accretion disks in outburst.             

For a Keplerian thin disk the mass and angular momentum
conservation equations give a non-linear diffusion equation for the
surface density  

\begin{equation}
\frac{\d \Sigma}{\d t} = \frac{3}{R} \frac{\d}{\d R}
\left[ R^{1/2} \frac{\d}{\d R} (\nu \Sigma R^{1/2}) \right]
\label{1}
\end{equation} 
(Frank et al. 1992), 
where $\nu$ is the kinematic viscosity which,  together with the
surface densities, can be related to the disk midplane temperatures 
$T_{\m{c}}$ through   
\begin{equation}  
\frac{4 \sigma}{3 \tau} T_{\m{c}}^{4} = \frac{9}{8} \nu \Sigma. 
\label{2}
\end{equation} 
$\tau=\kappa \Sigma$ is the vertically integrated optical
depth, and $\sigma$ is the Stefan-Boltzmann constant. 
For the viscosity we use the standard $\a$ prescription 
$\nu = \a c_{\m{s}} h$  (Shakura $\&$ Sunyaev 1973)  
where $c_{\m{s}} = k T_{\m{c}} / \mu m_{\m{p}}$ is the
local sound speed, $\mu$ the mean molecular weight,   
$h = c_{\m{s}}/\Omega_{\m{K}}$ the pressure scale height of the disk, 
and $\Omega_{\m{K}}$ the local Keplerian angular velocity of the disk. 
We use      
electron scattering opacities 
($\kappa_{\m{es}}\simeq 0.4$ cm$^2$ g$^{-1}$).  
We chose $\mu=0.6$ and $\a=0.1$ which is typical of the hot state
viscosities in the disk models of dwarf novae and 
soft X-ray transients.     

By setting $x=2 R^{1/2}$ and $ S =x \Sigma$, Eq.(\ref{1}) can be
written
in a simple form
\begin{equation}
\frac{\d S}{\d t} = \frac{12}{x^2} \frac{\d^2}{\d x^2} (\nu S) .
\label{3}    
\end{equation}
We divide the disk into 400 equally spaced grid points in $x$. 
This provides a better spatial resolution for the inner disk in
comparison to a model with the  same number of grid points equally spaced
in $R$. 

For a thin disk, the total disk luminosity is 
$\Ldisk= G M \Mdotin / 2 R_{\m{in}}$, and most of this emission
comes from the
inner disk, characterized by a disk black-body spectrum. 
Here, $\Mdotin$  is the mass inflow rate arriving at  the disk inner 
radius $\Rin$, and 
$M$ is the mass of the neutron star (NS). We take $M=1.4 M_{\odot}$
throughout the calculations.
The accretion luminosity from the NS surface,  
$L_*= G M \Mdot*/R_*$, determines the observed luminosity in the X-ray
band. The evolution of $\Mdotin(t)$ in the disk will be reflected in the 
accretion luminosity from the NS surface, depending on the fraction of 
matter accreted, $f= \Mdot*/\Mdotin$  
where $\Mdot*$ is the mass accretion 
rate onto the star.
We present three model calculations corresponding to 
different $f$ values (0.1, 0.5, 0.9). 

While the observed luminosity is expected to be powered by accretion 
onto the NS surface, the spectra during the enhanced X-ray emission 
of SGR 1900+14 can be fitted by a single power-law (Woods et al. 2001). 
A scattering
source, e.g. a hot corona, around the inner disk can significantly 
change the spectrum emitted from the neutron star surface and from the 
disk black-body spectrum into a power-law spectrum by means of
inverse Compton scatterings. If the source of the corona is fed by the
thermal instabilities at the surface (or inner rim) of the disk then 
the total luminosity remains constant for a given matter inflow rate 
and inner disk radius, while the spectrum 
may be modified from the input spectrum. 
Comparison of spectral models for emission from the NS surface 
or the disk  with  the observed $2-10$ keV band data may be misleading. 
We take the observed luminosity to
represent the total luminosity 
assuming that most of the X-ray flux from the source 
is emitted in the observation band (2-10 keV). 
For the model fits,  
we relate the model luminosities to the fluxes 
by $F_{\m{disk}}\sim (L_{\m{disk}} ~cos ~i)/(4 \pi d^2)$  
and $F_* \sim L_*/(4 \pi d^2)$ where
$d=14.5$ kpc is the distance of the source (Vrba et al. 2000). We   
set $ cos ~i= 0.8$ and 
neglected the small time delay for the
matter to travel from $R_{\m{in}}$ to $R_*$.   

\section{Results and Discussion }

The disk parameters for the model curves presented in Figures 1-3 
are given in Table 1. 
The lower and the upper model curves in the Figures correspond to the fluxes 
originating from the inner disk and from the NS surface respectively  
with $L_{*}= 2 (\Mdot*/R_*)(\Rin / \Mdotin)
L_{\m{disk}} = 2 f (R_{\m{in}}/R_*) L_{\m{disk}}$.   
For each of the three different $f$ values (0.1, 0.5, 0.9)  
$L_x >> \Ldisk$.   
Our models produce good fits to the
wide range of $f$. 
For each mass accretion ratio $f$, the quiescent luminosity gives the mass
inflow rate in the disk. The $\Rin$ values given in Table 1 are estimated 
Alfv$\acute e$n radii for these mass inflow rates, taking the dipole magnetic
moment $\mu =10^{30}$ G cm$^3$.   
These results strongly suggest a viscously
evolving disk origin for the observed post burst X-ray enhancement, 
but do not constrain $f$. 
In this range of $f$, rough estimates with the thin disk model give 
$m_I\simeq 19-20.5$ at the peak of the light curve, and 
 $m_I\simeq 26$, similar to the upper limits in the quiescent 
 phase (Vrba et al, 2000). The upper limits placed by the IR
observations
about 8 days after the giant flare, when the X-ray flux has decreased  
to about one percent of its peak level, are 
$m_J\ga ~22.8$ and $m_{K_s}\ga ~20.8$ (Kaplan et al. 2002). 
The IR expected light curve during the  
X-ray enhancement  and in quiescence will be presented  
in a separate work.
 
The energy given to the disk by the
giant flare could be written as 
$\delta E= \beta \dot{E} \Delta t \sim \beta 10^{44} $ ergs where
$\beta= \Bb+\Be$ is
the fraction of the total flare energy absorbed by the disk. 
Part of the inner disk matter  heated by the  energy $\Be \delta E$ 
can escape from
the system, while the remaining part is pushed back by $\Bb \delta E$ 
staying bound and piling up
at the inner rim of the disk. 
$\beta$ is expected to be around 
$\sim 2 \pi (2H_{\m{in}}) \Rin/ 4\pi \Rin^2 = H_{\m{in}}/\Rin \sim$
few $\times 10^{-3}$ for a thin disk 
with $\dot{M}\sim 10^{15-16}$ g s$^{-1}$ where $H_{\m{in}}$ is the
semi-thickness  of the 
disk at $\Rin$. This ratio is roughly constant throughout the disk (e.g.
Frank et al. 1992). 
The energy imparted by the flare to push back the inner disk matter is:
$\delta E_{\m{b}}\simeq (GM\delta M/2\Rin) [1-(\Rin/R_0)]$.  
 This is almost equal to the binding energy,  
since we find that $\Rin/R_0 \sim 1/3$ 
for the models given in Table 1. 
The energy used up pushing back the disk is a fraction of the 
estimated energy, absorbed by the disk, $\Bb < \beta$. It is in fact  
likely that a larger amount of matter escapes from the system, than 
the amount $\delta M$ that is pushed back but remains bound, with 
$\Be \sim (5-25) \Bb$.     

The maximum amount of mass that can escape from the inner
disk during a burst can be 
estimated as  $\delta M_{\m{loss}}\sim (2 \Rin /GM) \beta \delta E\simeq 
10^{23}$ g $R_{\m{in,8}} ~(\beta /10^{-3})$ where $R_{\m{in,8}}$ 
is the inner disk radius in units of $10^8$ cm.
 During the lifetime of an SGR ($\sim 10^4$ yrs) which  
has a giant burst per century, the total mass loss would be 
 $10^{25}$ g $R_{\m{in,8}} ~(\beta /10^{-3})$.

If the pulsed fraction remains the same ($\sim 0.1$) throughout
the enhanced X-ray flux phase as estimated by Woods at al. (2001) 
we expect a connection between the mass inflow rate and the pulsed X-ray 
emission. In our models, the luminosity from the
NS surface dominates the disk luminosity, and the   
pulsed fraction ${\cal F}\sim 0.1$ 
could be explained as the ratio 
of  the emission beamed by the mass flow geometry through the polar
caps to the isotropic emission from or near the NS surface.

The time evolution in our models is quite prompt, with a viscous time
scale  $t_{\nu} \sim R^2 /\nu \sim 10^3$ s, in agreement with the observed 
X-ray enhancement. Thompson et al. (2000)
estimate a viscous time scale of $\sim 10$ yrs 
for the reestablishment of the inner disk   
mainly because they use   
pre burst mass flow rate $\dot{M}\simeq 10^{15}$ g s$^{-1}$
in their estimate, 
instead of the appropriate post burst $\dot{M}$,     
which is three orders of magnitude higher.  
Thompson et al. also take 
$\alpha=0.01$ and estimate the post burst inner disk radius to be  
$R_0= 10^{10}$ cm. In our calculations, 
$\alpha=0.1$, typical of the outburst (hot)
states of the soft X-ray transient and dwarf nova disk models. 
The post burst pile up position $R_0 \sim 10^{9}$ cm  
in our models corresponds to the short viscous time scale.   
For smaller burst energies  
$(10^{41-42}$ ergs), the inner disk matter is pushed out 
to correspondingly smaller radii $R_{0}$, 
and  $t_{\nu}$ could be as small as 
a few seconds.      

The enhanced mass inflow rate can modify the spin evolution significantly 
especially around the peak of the X-ray light curve when $L_{\m{x}}$ is 
about $\sim 700$ times higher than its quiescent level. A similar increase in 
$\dot{M}_{\m{in}}$ corresponds to a decrease in the Alfv$\acute e$n  radius 
$r_{\m{A}}$ by a factor
$\sim 6$ (or less if the field lines are compressed by rapid accretion). 
If the fastness parameter $\omega=\Omega_*/\Omega_{\m{K}}(r_{\m{A}})$
drops below unity a spin up is expected to prevail until $\omega$ has reduced
back. If the system does not enter the spin-up phase or 
remains mostly in the spin-down phase in the high mass inflow regime, a
sharp increase in the spin down torque could abruptly reduce the spin frequency.   
The changes in the spin evolution depends not only on the variations of the 
$\dot{M}_{\m{in}}$ and $\omega$ but also on how close the system is to 
rotational equilibrium in quiescence and on the inner disk 
structure (likely to be somewhat different from a thin disk geometry) 
during the X-ray enhancement phase. Due to these uncertainties  
it is hard to make a reliable estimate of the spin evolution  during the 
short term unsteady phase of these systems. A detailed examination  of the
possible post-burst spin evolution considering different 
accretion geometries will be addressed in future work.

Observations of four bursts from SGR 1900+14, including the August 27
giant flare and 3 smaller events,  extending three orders of
magnitude in flare fluence were studied by Lenters et al. 
(2003, see especially their Fig. 13). Their study reveals that the ratio
of the fluence of the enhanced X-ray emission $\dEx$ to the fluence of the 
preceding burst energy $\dEb$ is $\sim 0.02$, and remains constant from burst
to burst. In our models, this ratio can be written as 
$\gamma= \dEx/\dEb \simeq 2 \Bb f (\Rin/R_*)$ where both 
$\Bb$ and $\Rin$ represent the pre burst inner disk conditions. 
$\Bb$ depends on the disk geometry and is very likely to be similar prior to 
the different bursts of SGR 1900+14. Our models with a constant $f$ along the
X-ray enhancement phase fits well to the data  indicating that 
$f$  remains constant along this phase. Since the X-ray enhancements 
following the three other smaller events  trace accretion rates 
that were encountered along the decaying tail of the 
post giant flare enhancement, a similar $f$ 
must be operating throughout the smaller enhancements following the three 
events. The remaining variable 
$\Rin$  depends on the pre burst $\Mdotin$. An order of magnitude change in 
$\Mdotin$ causes a change in $\Rin$ by a factor $\la ~2$. So, based on our
model results, it is understandable 
that the ratio $\gamma$ remains constant  within a factor
$\sim 2$ for different bursts of a particular SGR, consistent with the
observations of SGR 1900+14. $\gamma$ may vary from source to source depending
on the preburst inner disk conditions. 
                
\section{Conclusion}

We have shown that   the X-ray flux curve
following the 1998 August 27 
giant flare of SGR 1900+14 
can be accounted for by the enhanced accretion onto the neutron star
surface due to the  relaxation of the
disk, starting from new initial conditions with the inner disk pushed back  
by a plausible fraction of the flare energy. 
For our disk models, the ratio of the fluence of the 
X-ray enhancement to the preceding burst energy remains roughly
constant for bursts of a given SGR with similar preburst mass inflow rates, 
in agreement with the burst and enhancement observations of 
SGR 1900+14 (Lenters et al. 2003).
This ratio can vary for different SGRs indicating their different 
inner disk conditions.

\acknowledgments

We thank Ersin G\"{o}\u{g}\"{u}\c{s} and Peter Woods for discussions, 
and the referee for usefulcomments. 
We acknowledge support from the High Energy Astrophysics Research Group
TBAG-\c{C}-4 of T{\"U}B{\.I}TAK and from the Astrophysics and Space
Forum at Sabanc{\i} University.
MAA acknowledges support from the Turkish Academy of Sciences.


\clearpage
\begin{figure}
\plotone{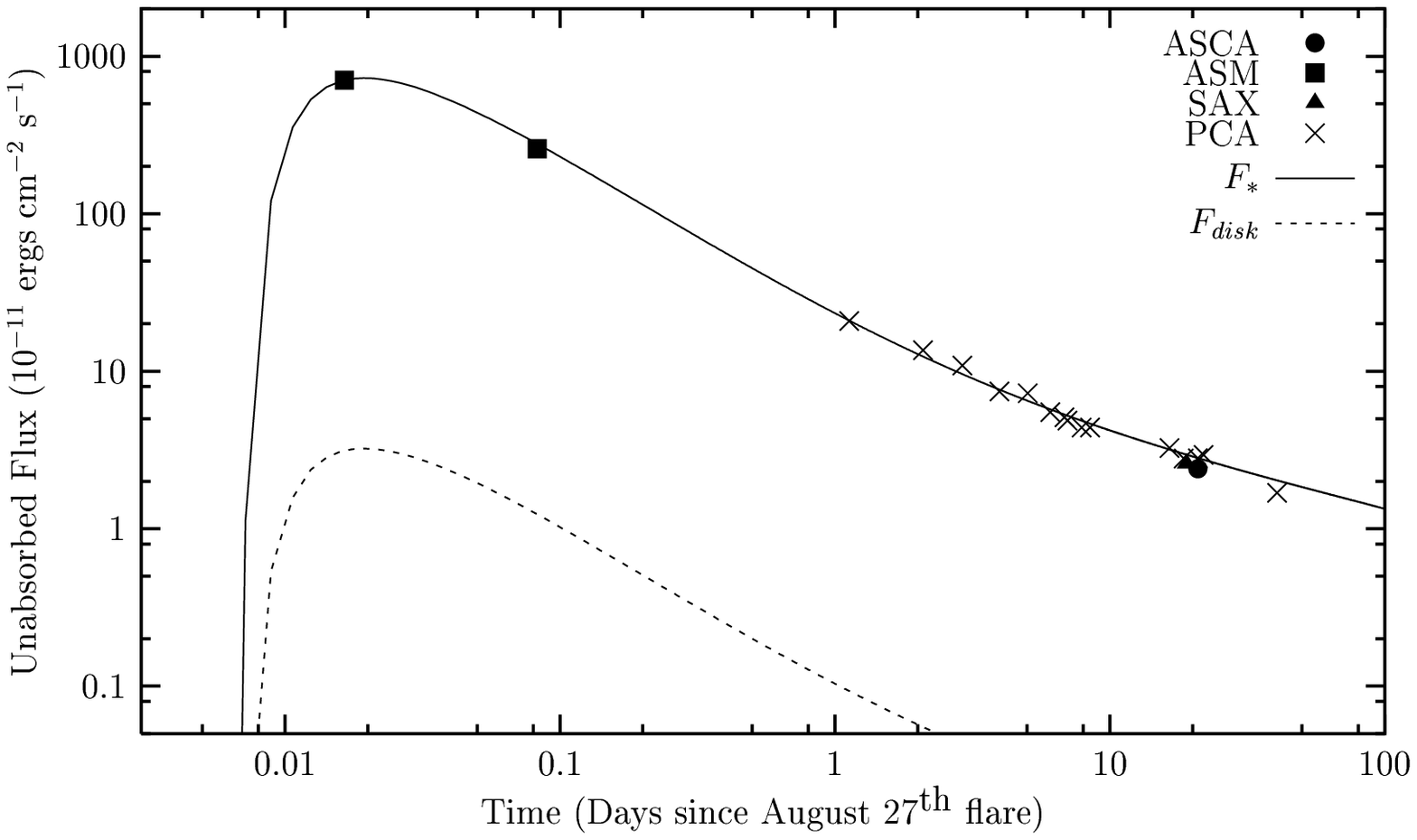}
\caption{The data points (RXTE/ASM, RXTE/PCA, BeppoSAX and ASCA
measurements) was taken from Woods et al.(2001) (see the text
for the details and uncertainties of the measurements). 
The upper curve is the model flux from the surface of the neutron star
and the lower curve is the model disk
flux. For this illustrative model (MODEL 1), $f\simeq
0.1$. The parameters of Models 1$-$3 are presented in Table 1.}

\end{figure}

\clearpage
\begin{figure}
\plotone{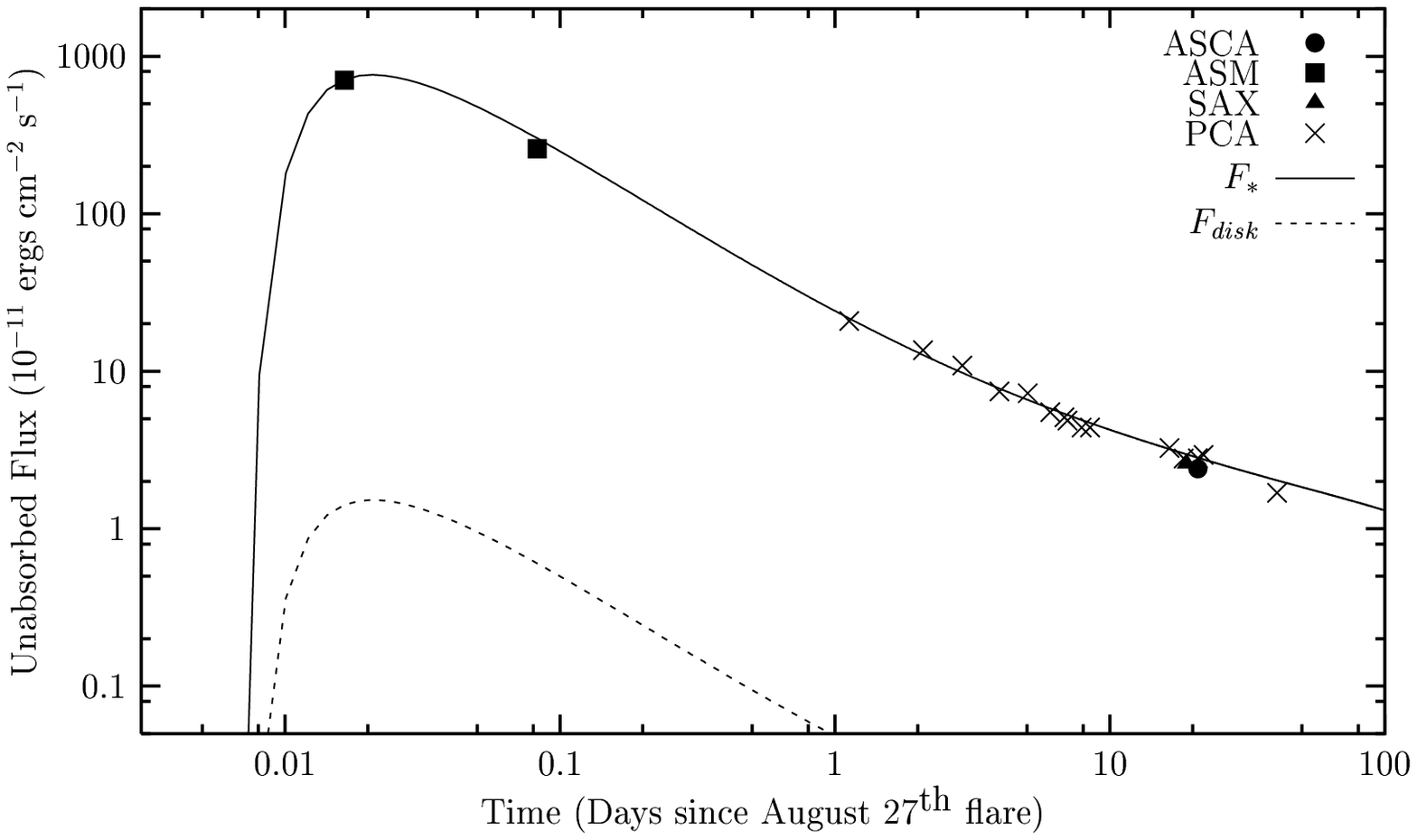}
\caption{Same as Fig. 1 except $f=0.5$.} 
\end{figure}

\clearpage
\begin{figure}
\plotone{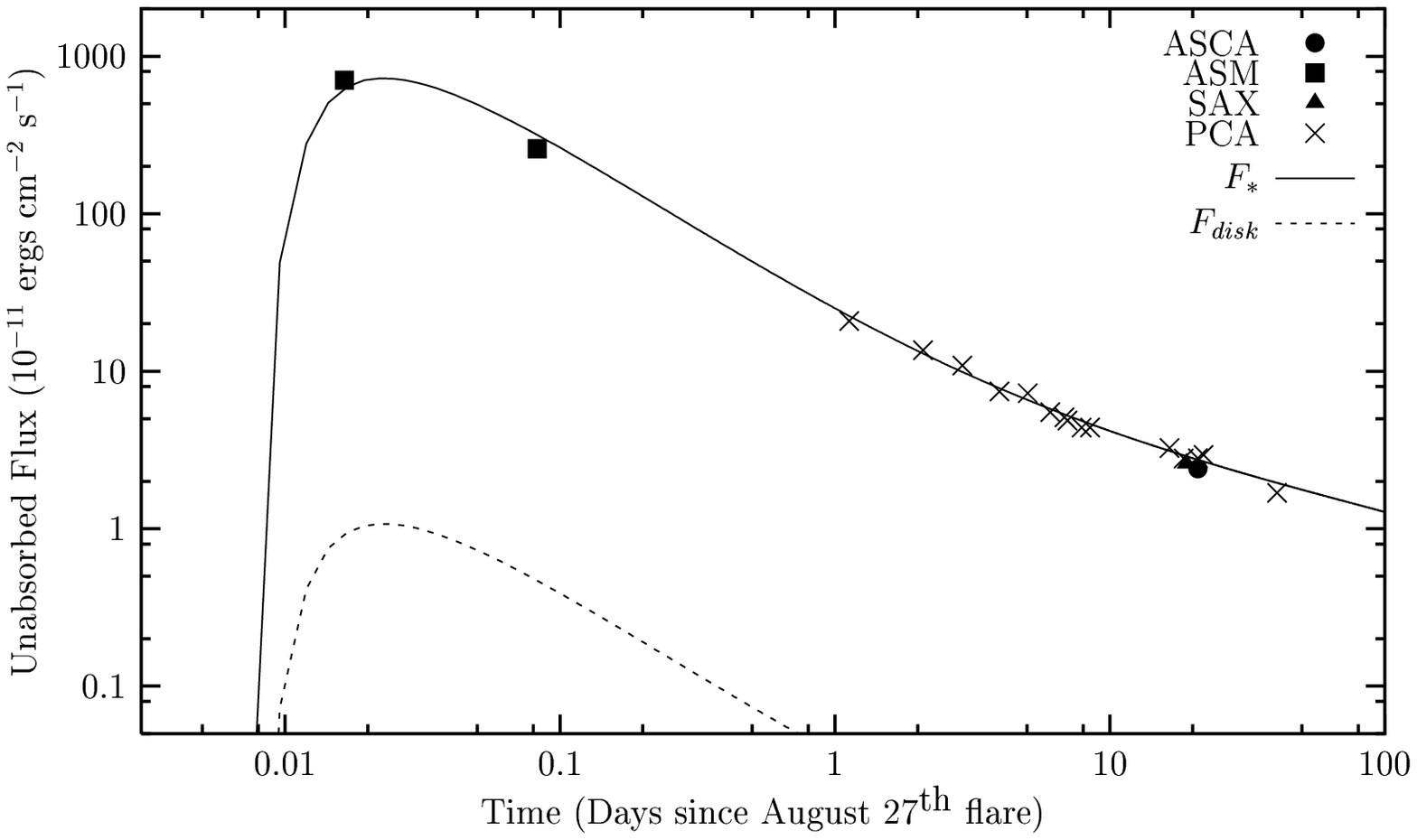}
\caption{Same as Fig. 1 except $f=0.9$.}
\end{figure}

\clearpage

\begin{table}
\begin{center}
\caption{Model parameters for the flux evolution presented in Fig. 1. 
In all model calculations, the viscosity parameter $\alpha =0.1$, 
the mean molecular weight $\mu =0.6$ and electron scattering opacities 
are used}

\begin{tabular}{l|c|c|c}
\tableline\tableline
&{MODEL 1}&{MODEL 2}&{MODEL 3}
\\
\hline
$\Sigma_{\m{max}}$ (g cm$^{-2}$)&$9.6\times 10^4$&$3.0\times 10^4$ 
&$2.1\times 10^4$   \\
\hline
Gaussian width (cm)&$2.4\times 10^7$ &$2.2\times 10^7$ 
&$2.2\times 10^7$\\
\hline
$\Sigma_{0}/\Sigma_{\m{max}}$&0.012&0.020&0.022\\
\hline
$R_0$ (cm)&$1.8\times 10^9$& $1.1\times 10^9$ 
&$9.4\times 10^8$ \\
\hline
$R_{\m{in}}$ (cm)&$6.0\times 10^8$ &$4.0\times 10^8$
&$3.0\times 10^8$  \\
\hline
$R_{\m{out}}$ (cm)&$1.0\times 10^{11}$ &$1.0\times 10^{11}$ 
&$1.0\times 10^{11}$ \\
\hline
$f$ &0.1&0.5&0.9\\
\hline
estimated $\Bb$&$2\times 10^{-4}$&$5\times 10^{-5}$&
$4\times 10^{-5}$\\
\hline
$\delta M$ (g)&$2\times 10^{23}$&$3.5\times 10^{22}$&
$2\times 10^{22}$\\
\tableline

\end{tabular}
\end{center}
\end{table}

\end{document}